\documentstyle[referee]{laa}

\begin{document}
\thesaurus{07  
	   (07.09.1; 
	    07.13.1;  
	   )}

\title{Electromagnetic Radiation and Motion of Really Shaped Particle}
\author{J.~Kla\v{c}ka}
\institute{Institute of Astronomy,
   Faculty for Mathematics and Physics, Comenius University \\
   Mlynsk\'{a} dolina, 842~48 Bratislava, Slovak Republic}
\date{}
\maketitle

\begin{abstract}
Relativistically covariant form of equation of motion for real particle
(neutral in charge) under the action of electromagnetic radiation is derived.
Various formulations of the equation of motion in the proper frame of reference
of the particle are used.

Main attention is devoted to the reformulation of the equation of motion
in the general frame of reference, e. g., in the frame of reference of the
source of electromagnetic radiation. This is the crucial form of equation
of motion in applying it to motion of particles (cosmic dust, asteroids, ...)
in the Universe if electromagnetic radiation acts on the particles.

General relativistic equation of motion is presented.

\keywords{relativity theory, cosmic dust, asteroids}

\end{abstract}

\section{Introduction}
Covariant form of equation of motion for a particle is required if one wants
to be sure that the motion of the particle is correctly described. This is
the crucial fact of relativity theory, firstly understood by Minkowski.
Thus, if we want to derive equation of motion of a particle under the
action of electromagnetic radiation, we have to bear in mind this crucial
requirement of relativity theory. Bearing this requirement in mind,
Kla\v{c}ka (2000a) has derived covariant form of equation of motion for
really shaped dust particle under the action of electromagnetic radiation.
However, even one year ago the general opinion was that its significance is
negligible. Two conferences (US-European Celestial Mechanics Workshop,
Poznan, Poland, July 2000; Light Scattering by Nonpsherical Particles:
halifax Contributions, Halifax, Canada, August/September 2000) have changed
the opinion of several astronomers. Although the original papers
Kla\v{c}ka (2000a, 2000b, 2000c) has not succeeded in journal publication,
one consequence of Kla\v{c}ka (2000a) -- derivation to the first order
in $v/c$ (higher orders are neglected; $\vec{v}$ is velocity
of the particle, $c$ is the speed of light) -- with application to real system
in the Solar System is published (Kla\v{c}ka and Kocifaj 2001); accuracy
to the first order in $v/c$ may be sufficient in many applications in practice.

Several authors have tried to derive equation of motion for really shaped
dust particle under the action of electromagnetic radiation during the last
several months. However, they have not succeeded, and, moreover,
the paper Kla\v{c}ka (2000a) has not help them in
correct understading of the physics of the phenomenon.
(According to their opinion publication in international journal
is required, not only web-form.)
This motivates to
derive covariant form of equation of motion for particle under the action
of electromagnetic radiation in the way which will stress the crucial steps.
Thus, we will take equation of motion in the proper frame of reference
as the initial point of our paper. We will rewrite it to covariant form,
in this paper. We will use the results of Kla\v{c}ka (2000a, 2001). However,
derivations will be more simple and a little generalized results will be
obtained. Moreover,
general relativistic equation of motion will be presented.

\section{Generalized special Lorentz transformation}
By the term ``stationary frame of reference'' we shall mean a frame of reference
in which particle moves with a velocity vector $\vec{v} = \vec{v} (t)$.
The physical quantities measured in the stationary frame of reference
will be denoted by unprimed symbols.

The term ``stationary particle'' will denote particle which does
not move in a given inertial frame of reference.
Primed quantities will denote quantities measured in the
proper reference frame of the particle.

Our situation corresponds to the fact that we know equation of motion
of the particle in its proper frame of reference. We want to derive
equation of motion for the particle in the
stationary frame of reference.

We have to use generalized special Lorentz transformation for the purpose of
making transformation from proper frame of reference to stationary frame
of reference.

If we have a four-vector $A^{\mu} = ( A^{0}, \vec{A} )$, where
$A^{0}$ is its time component and $\vec{A}$ is its spatial component,
generalized special Lorentz transformation yields
\begin{eqnarray}\label{1}
A^{0 '}  &=& \gamma ~ ( A^{0} ~-~ \vec{v} \cdot \vec{A} / c ) ~,
\nonumber \\
\vec{A} ' &=& \vec{A} ~+~ [ ( \gamma ~-~ 1 ) ~ \vec{v} \cdot \vec{A}  /
	      \vec{v} ^{2} ~-~ \gamma ~ A^{0} / c ] ~ \vec{v}  ~.
\end{eqnarray}
The inverse generalized special Lorentz transformation is
\begin{eqnarray}\label{2}
A^{0}  &=& \gamma ~ ( A^{0 '} ~+~ \vec{v} \cdot \vec{A} ' / c ) ~,
\nonumber \\
\vec{A}  &=& \vec{A} ' ~+~ [ ( \gamma ~-~ 1 ) ~ \vec{v} \cdot \vec{A} ' /
	      \vec{v} ^{2} ~+~ \gamma ~ A^{0 '} / c ] ~ \vec{v}  ~.
\end{eqnarray}
The $\gamma$ factor is given by the well-known relation
\begin{equation}\label{3}
\gamma = 1 / \sqrt{1~-~\vec{v} ^{2} / c ^{2} } ~.
\end{equation}

As for four-vectors we can immediately introduce four-momentum:
\begin{equation}\label{4}
p^{\mu} \equiv ( p^{0}, \vec{p} ) = ( E / c, \vec{p} ) ~.
\end{equation}

\section{First derivation}

\subsection{Proper reference frame of the particle -- stationary particle}
As it was already stated,
primed quantities will denote quantities measured in the
proper reference frame of the particle.

Equation of motion of the particle in its proper frame of refrence
is taken in the form
\begin{eqnarray}\label{5}
\frac{d~ E'}{d~ \tau} &=& 0 ~,
\nonumber \\
\frac{d~ \vec{p'}}{d~ \tau} &=& \frac{1}{c} ~ S'~A'~ \left \{
	     Q_{R} ' ~ \vec{S}_{i} ' ~+~ Q_{1} '
	     ~ \vec{e}_{1} ' ~+~ Q_{2} '
	     ~ \vec{e}_{2} ' \right \} ~,
\end{eqnarray}
where $E'$ is particle's energy, $\vec{p'}$ its momentum, $\tau$ is proper
time, $c$ is the speed of light, $S'$ is the flux density of radiation energy
(energy flow through unit area perpendicular to the ray per unit time),
$A'$ is geometrical cross-section of a sphere of volume equal to
the volume of the particle, $Q'_{R}$, $Q'_{1}$ and $Q'_{2}$ represent
effective factors of radiation pressure,
unit vector $\vec{S}_{i} '$ is directed along the path of the incident
radiation (it is supposed that beam of photons propagate
in parallel lines) and its orientation corresponds to the orientation
of light propagation, the unit vectors $\vec{S}_{i} '$, $\vec{e}_{1} '$
and $\vec{e}_{2} '$ used on the RHS of Eqs. (5) form an orthogonal basis.

Eqs. (5) describe equation of motion of the particle in the proper
frame of reference due to its interaction with electromagnetic radiation.
(It is supposed that the energy $E'$ of the particle is unchanged:
the energy of the incoming radiation equals
to the energy of the outgoing radiation, per unit time.)

\subsection{Transformation of the required quantities}

Our aim is to derive equation of motion for the particle in the
stationary frame of reference. We will use the fact that we know
this equation in the proper frame of reference -- see Eqs. (5).
We have to use generalized special Lorentz transformation for the purpose of
making transformation from proper frame of reference to stationary frame
of reference.
We have to rewrite Eqs. (5) into covariant form. We have to transform
the corresponding primed quantities according to the generalized special
Lorentz transformation, as it was described in the section 2.

As for the transformation of the LHS of Eqs. (5), there is no problem:
Eqs. (2) and Eq. (4) yield
\begin{eqnarray}\label{6}
\frac{d~ E}{d~ \tau}   &=& \gamma ~
		    \vec{v} ~ \cdot \frac{d~ \vec{p'}}{d~ \tau}  ~,
\nonumber \\
\frac{d~ \vec{p}}{d~ \tau}   &=& \frac{d~ \vec{p'}}{d~ \tau}  ~+~
	   \left ( \gamma ~-~ 1 \right ) ~
	   \left ( \vec{v} \cdot \frac{d~ \vec{p'}}{d~ \tau} \right )
	   ~ \frac{\vec{v}}{\vec{v} ^{2}}  ~.
\end{eqnarray}

Using the fact that
$p^{\mu} = ( h ~\nu , h ~\nu ~ \hat{\vec{S}_{i}} )$ for the incident photon,
Lorentz transformation yields
\begin{eqnarray}\label{7}
\nu ' &=& \nu  ~w ~,
\nonumber \\
\vec{S}_{i} ' &=& \frac{1}{w}  ~ \left \{ \vec{S}_{i} ~+~
		 \left [ \left ( \gamma ~-~ 1 \right ) ~
		 \vec{v} \cdot \vec{S}_{i}  /
		 \vec{v} ^{2} ~-~ \gamma / c \right ] ~ \vec{v} \right \} ~,
\end{eqnarray}
where abbreviation
\begin{equation}\label{8}
w \equiv \gamma ~ ( 1 ~-~ \vec{v} \cdot \vec{S}_{i} / c )
\end{equation}
is used.

As for the transformation of the flux density of radiation energy, we
use derivation presented in Kla\v{c}ka (1992).
We can write for monochromatic radiation
\begin{equation}\label{9}
S' = n' ~h~ \nu ' ~ c ~; ~~~ S = n ~h~ \nu  ~ c ~,
\end{equation}
where $n$ and $n'$ are concentrations of photons in corresponding
reference frames.
The continuity equation
\begin{equation}\label{10}
\partial _{\mu} ~ j^{\mu} = 0 ~, ~~~j^{\mu} = ( c~n, c~n~ \vec{S}_{i} ) ~,
\end{equation}
holds for the four-vector of the current density $j^{\mu}$.
Using Eqs. (1) one can easily obtain
\begin{equation}\label{11}
n' = w ~ n~.
\end{equation}
On the basis of Eqs. (7), (9) and (11) we finally obtain
\begin{equation}\label{12}
S' = w^{2} ~ S ~.
\end{equation}

How to find transformations for unit vectors $\vec{e}_{1} '$
and $\vec{e}_{2} '$? The crucial point is what physics do these
unit vectors describe. They (also, together with unit vector $\vec{S}_{i} '$)
describe the directions of propagation of the light after interaction
with the particle. Thus, the abberation of light also exists for each of
these unit vectors. Considerations analogous to those for vector $\vec{S}_{i} '$
immediately yield:
\begin{eqnarray}\label{13}
\vec{e}_{j} ' &=& \frac{1}{w_{j}}  ~ \left \{ \vec{e}_{j} ~+~
		 \left [ \left ( \gamma ~-~ 1 \right ) ~
		 \vec{v} \cdot \vec{e}_{j}  /
		 \vec{v} ^{2} ~-~ \gamma / c \right ] ~ \vec{v} \right \} ~,
\nonumber \\
w_{j} &\equiv& \gamma ~ ( 1 ~-~ \vec{v} \cdot \vec{e}_{j} / c ) ~~, ~~~j = 1, 2 ~.
\end{eqnarray}

On the basis of Kla\v{c}ka (1992) we know that the quantity
$w^{2} ~ S ~ A'$ is scalar.

\subsection{Covariant form of equation of motion}

Putting RHS of Eqs. (5) into Eqs. (6), and, using Eqs. (7), (12) and (13),
one can obtain
\begin{eqnarray}\label{14}
\frac{d ~E}{d~ \tau} &=&  \frac{w^{2} ~ S ~A'}{c} ~ \left \{
	  Q_{R} ' ~  \left ( c ~\frac{1}{w} ~-~  \gamma ~c \right ) ~+~
	  \sum_{j=1}^{2} ~Q_{j} ' ~ \left ( c ~\frac{1}{w_{j}}
	  ~-~ \gamma ~c \right ) \right \} ~,
\nonumber \\
\frac{d ~\vec{p}}{d~ \tau} &=& \frac{w^{2} ~ S ~A'}{c^{2}} ~ \left \{
      Q_{R} ' ~ \left ( c ~ \frac{\vec{S}_{i}}{w} ~-~ \gamma ~ \vec{v} \right )
     ~+~ \sum_{j=1}^{2} ~ Q_{j} ' ~ \left ( c~
				    \frac{\vec{e}_{j}}{w_{j}}
				    ~-~ \gamma ~ \vec{v}  \right ) \right \}  ~.
\end{eqnarray}

Comparisons of Eqs. (2), (7) and (13) yield that we have these three
four-vectors:
\begin{eqnarray}\label{15}
b_{i}^{\mu} &=& ( 1 / w , \vec{S}_{i} / w ) ~,
\nonumber \\
b_{j}^{\mu} &=& ( 1 / w_{j} , \vec{e}_{j} / w_{j} ) ~, ~~ j= 1, 2 ~.
\end{eqnarray}

Considering the fact that $p^{\mu} = ( E ~/~ c,~ \vec{p} )$ is four-momentum
of the particle of mass $m$
\begin{equation}\label{16}
p^{\mu} = m~ u^{\mu} ~,
\end{equation}
four-vector of the world-velocity of the particle is
\begin{equation}\label{17}
u^{\mu} = ( \gamma ~c, \gamma ~ \vec{v} )
\end{equation}
and that other four-vectors are defined by Eqs. (15),
one easily obtains that
Eqs. (14) may be rewritten in terms of four-vectors:
\begin{equation}\label{18}
\frac{d ~p^{\mu}}{d~ \tau} =  \frac{w^{2} ~ S ~A'}{c^{2}} ~ \left \{
      Q_{R} ' ~ \left ( c ~ b_{i}^{\mu} ~-~ u^{\mu} \right )
     ~+~ \sum_{j=1}^{2} ~ Q_{j} ' ~ \left ( c~ b_{j}^{\mu} ~-~ u^{\mu} \right )
				   \right \}  ~.
\end{equation}

Eq. (18) is covariant form of equation of motion for the particle
moving in the field of electromagnetic radiation. Since we have not
transformed the quantities $Q'_{R}$, $Q'_{1}$ and $Q'_{2}$, these quantities
are scalars, invariants of the generalized special Lorentz transformation
(mass $m$ of the particle is also scalar, invariant).

It can be easily verified that Eq. (18) yields $d ~m / d~ \tau =$ 0.

\section{Second derivation}

\subsection{Proper reference frame of the particle -- stationary particle}
Again, primed quantities will denote quantities measured in the
proper reference frame of the particle.

Let the equation of motion of a particle in the electromagnetic radiation
field is expressed in the form
\begin{eqnarray}\label{19}
\frac{d \vec{p'}}{d \tau} &=& \frac{1}{c} ~S'~\left ( C'~\vec{S'_{i}}
			   \right )
\nonumber \\
\frac{d E'}{d \tau} &=& 0 ~;
\end{eqnarray}
$S'$ is the flux density of the radiation
energy, $C'$ is pressure cross section 3 $\times$ 3 matrix,
$\vec{S'_{i}}$ is unit vector of the incident radiation, $\tau$ is
proper time, $c$ is the speed of light.

We are interested in deriving equation of motion of the particle in the
rest frame of the source: the particle moves with instantaneous velocity
$\vec{v}$ with respect to the source, the unit vector of the incident
radiation is $\vec{S_{i}}$ and other physical quantities measured
in the rest frame of the source are also unprimed.

\subsection{Reformulation of the initial equation of motion}
We will be inspired by ideas presented in Kla\v{c}ka (2001) -- the ideas
will be a little generalized.

Let the components of the pressure cross section $C'$ 3 $\times$ 3 matrix
are given in a basis of orthonormal vectors
$\vec{e}'_{b1}$, $\vec{e}'_{b2}$, $\vec{e}'_{b3}$.
We may write, then
\begin{equation}\label{20}
\vec{S'_{i}} = \sum_{k=1}^{3} ~ s'_{k} \vec{e}'_{bk} ~.
\end{equation}

Using Eq. (20) we have
\begin{equation}\label{21}
C'~\vec{S'_{i}} = \sum_{k=1}^{3} ~\left (  \sum_{l=1}^{3} ~ C'_{kl} ~ s'_{l}
		  \right ) ~ \vec{e}'_{bk} ~.
\end{equation}

On the basis of Eq. (21) we can rewrite the first of Eqs. (19) to the form
\begin{equation}\label{22}
\frac{d \vec{p'}}{d \tau} = \frac{1}{c} ~S'~ \sum_{k=1}^{3} ~
\left ( \vec{e}_{bk} ^{'T} ~C'~\vec{S'_{i}} \right )
~\vec{e}'_{bk} ~.
\end{equation}

\subsection{Covariant form of equation of motion}

Comparison of Eq. (22) with Eq. (5) enables
immediately write covariant form of equation of motion, on the basis
of Eq. (18):
\begin{equation}\label{23}
\frac{d p^{\mu}}{d \tau} = \frac{w^{2} ~S}{c^{2}} ~
			   \sum_{j=1}^{3} ~
\left ( \vec{e}^{'T}_{bj} ~C'~\vec{S'_{i}} \right )~
\left ( c ~b_{bj}^{\mu} ~-~ u^{\mu} \right )   ~,
\end{equation}
where $p^{\mu} = m~u^{\mu}$, $u^{\mu} = (\gamma ~c, ~\gamma ~\vec{v} )$,
$w = \gamma ~( 1 ~-~\vec{v} \cdot \vec{S_{i}} ~/ ~c )$,
$b_{bj}^{\mu} = ( 1~/ ~w_{bj}, \vec{e}_{bj}~/ ~w_{bj} )$,
$w_{bj} = \gamma ~( 1 ~-~\vec{v} \cdot \vec{e}_{bj} ~/ ~c )$
(relations analogous to those given by Eqs. (13) hold for
$\vec{e}_{bj}$ and $\vec{e}'_{bj}$, $j=$ 1 to 3).

\subsection{Is everything correct?}

Eq. (23) reduces to Eq. (7) in Kla\v{c}ka (2001) for the case
$\vec{e}'_{b1} = \vec{e}'_{1}$,
$\vec{e}'_{b2} = \vec{e}'_{2}$,
$\vec{e}'_{b3} = \vec{S'_{i}}$. These substitutions correspond to Eq. (18),
also.

It is important to show that the final equation of motion does not
depend on the choice of the set of orthonormal basis vectors. If this would
not be true, then the results represented by Eqs. (5) and (18), or, by
Eqs. (22) and (23), are physically incorrect.

Thus, let us consider two sets of orthonormal basis vectors:
$\vec{e}'_{b1}$, $\vec{e}'_{b2}$, $\vec{e}'_{b3}$, and,
$\vec{e}'_{1}$, $\vec{e}'_{2}$, $\vec{e}'_{3}$ (we may take
$\vec{e}'_{3} = \vec{S'_{i}}$, as an example).

First of all, we can immediately write
\begin{equation}\label{24}
C'~\vec{S'_{i}} = \sum_{k=1}^{3} ~
       \left ( \vec{e}_{bk} ^{'T} ~C'~\vec{S'_{i}} \right ) ~\vec{e}'_{bk}
		= \sum_{k=1}^{3} ~
       \left ( \vec{e}_{k} ^{'T} ~C'~\vec{S'_{i}} \right ) ~\vec{e}'_{k} ~,
\end{equation}
since the decomposition of a vector into a basis of orthonormal vectors is unique.
This means that equation of motion in the proper frame of the particle does
not depend on the basis vectors -- this corresponds to correct physics.

Now, let us show that equation of motion given by Eq. (23) is also independent
on the basis of orthonormal vectors. On the basis of definition of four-vectors
\begin{eqnarray}\label{25}
b_{bj}^{\mu} &=& ( 1 / w_{bj}, ~ \vec{e}_{bj} / w_{bj} ) ~,
w_{bj} \equiv \gamma ~ ( 1 ~-~ \vec{v} \cdot \vec{e}_{bj} / c ) ~~,
\nonumber \\
b_{bj}^{' ~\mu} &=& ( 1, \vec{e}_{bj} ')  ~~...~proper frame
						~~, ~~~j = 1, 2, 3
\end{eqnarray}
and generalized special Lorentz transformation represented by Eq. (2), we
can write
\begin{eqnarray}\label{26}
b_{bj}^{0} &\equiv& \frac{1}{w_{bj}} =
       \gamma ~ ( 1 ~+~ \vec{v} \cdot \vec{e}_{bj} ' / c ) ~,
\nonumber \\
\vec{b}_{bj} &\equiv& \frac{\vec{e}_{bj}}{w_{bj}} =
	     \vec{e}_{bj} ' ~+~ \left [ \left ( \gamma ~-~ 1 \right ) ~
	     \frac{\vec{v} \cdot \vec{e}_{bj} '}{ \vec{v} ^{2}}
	     ~+~ \frac{\gamma}{c} \right ] ~ \vec{v} ~~,~~~j = 1, 2, 3 ~.
\end{eqnarray}

Let us calculate the important part of the RHS of Eq. (23),
$c ~b_{bj}^{\mu} ~-~ u^{\mu}$ -- Eqs. (26) are used:
\begin{eqnarray}\label{27}
c~b_{bj}^{0} ~-~ u^{0} &\equiv& c~b_{bj}^{0} ~-~ \gamma ~c  =
       \gamma ~ ( \vec{v} \cdot \vec{e}_{bj} ') ~,
\nonumber \\
c~ \vec{b}_{bj} ~-~ \vec{u} &\equiv& c~\vec{b}_{bj} ~-~ \gamma ~\vec{v}  =
	     c~ \vec{e}_{bj} ' ~+~ \left ( \gamma ~-~ 1 \right ) ~
	     \frac{c~ \vec{v}}{\vec{v} ^{2}} ~
	     \left ( \vec{v} \cdot \vec{e}_{bj} ' \right ) ~~,~~~j = 1, 2, 3 ~.
\end{eqnarray}
Putting Eqs. (27) into Eqs. (23), we obtain
\begin{eqnarray}\label{28}
\frac{d E}{d \tau} &=& \frac{w^{2} ~S}{c} ~ \gamma ~\vec{v} \cdot \vec{X} ~,
\nonumber \\
\frac{d \vec{p}}{d \tau} &=& \frac{w^{2} ~S}{c^{2}} ~ \left \{ c~ \vec{X}
	   ~+~ \left ( \gamma ~-~ 1 \right ) ~
	   \frac{c~ \vec{v}}{\vec{v} ^{2}} ~ \left ( \vec{v} \cdot \vec{X}
	   \right )~ \right \} ~,
\nonumber \\
\vec{X} &\equiv& \sum_{j=1}^{3} ~ \left (
	   \vec{e}^{'T}_{bj} ~C'~\vec{S'_{i}} \right ) ~ \vec{e}_{bj} ' ~.
\end{eqnarray}
On the basis of Eq. (24) and Eqs. (28) we can conclude that equation of
motion (Eq. (23)) does not depend on the chosen orthonormal basis of vectors.
This is required by physics.

\section{General relativity}

The generally covariant equation of motion can be immediately written on the
basis of Eq. (23):
\begin{equation}\label{29}
\frac{D p^{\mu}}{d \tau} = \frac{w^{2} ~S}{c^{2}} ~
			   \sum_{j=1}^{3} ~
\left ( \vec{e}^{'T}_{bj} ~C'~\vec{S'_{i}} \right )~
\left ( c ~b_{bj}^{\mu} ~-~ u^{\mu} \right )   ~,
\end{equation}
where the operator $D ~/~d \tau$ is the "total" covariant derivative
in the general relativistic sense and includes gravitational effects.

\section{Conclusion}

We have derived equation of motion for real, arbitrarily
shaped particle under the action
of electromagnetic radiation. It is supposed that equation of motion is
represented by Eqs. (5) in the proper frame of reference of the
particle, or, that interaction between the particle and electromagnetic
radiation is described by radiation pressure cross section $C'$ 3 $\times$ 3
matrix.
The final covariant form is represented by Eq. (18), or, by Eq. (23), or,
by Eq. (29).

Within the accuracy to the first order in $\vec{v} / c$, Eq. (18) yields
\begin{eqnarray}\label{30}
\frac{d~ \vec{v}}{d ~t} &=&  \frac{S ~A'}{m~c} ~ \left \{ Q_{R} ' ~ \left [
		 \left ( 1~-~ \vec{v} \cdot \hat{\vec{S}_{i}} / c \right ) ~
		 \hat{\vec{S}_{i}} ~-~ \vec{v} / c \right ] ~+~ \right .
\nonumber \\
& &  \left .  \sum_{j=1}^{2} ~Q_{j} ' ~\left [  \left ( 1~-~ 2~
	      \vec{v} \cdot \hat{\vec{S}_{i}} / c ~+~
	      \vec{v} \cdot \hat{\vec{e}_{j}} / c \right ) ~ \hat{\vec{e}_{j}}
	      ~-~ \vec{v} / c \right ] \right \} ~.
\end{eqnarray}
As for practical applications, the terms $v/c$ standing at $Q_{j}'$
are negligible for majority of real particles.
(We want to stress that values
of $Q'-$coefficients depend on particle's orientation with respect to the
incident radiation -- their values are time dependent.)

Application to larger bodies, e. g., asteroids, may be found in
Kla\v{c}ka (2000c).

\vspace*{0.5cm}

{\bf Acknowledgements:} This work was supported by the VEGA grant
No.1/7067/20.

\end{document}